\documentstyle [12pt]{article}
\large

\newcommand{\1}{\begin{equation}Q^{\mu}(z)=q^{\mu}-ip^{\mu}lnz+i\sum_{n\neq 0}
\frac{\alpha^{\mu}_{n}}{n}z^{-n}\end{equation}}

\newcommand{\2}{\begin{equation}U^{r}(z)=:e^{ir\cdot Q(z)}:\end{equation}}

\newcommand{\5}{\begin{eqnarray}
U^{\{r,r_{i}\}}_{\{(n_{i})\}}(z) & = & :\prod_{i}\lim_{z_{i}\rightarrow z}
\frac{1}{n_{i}!}\frac{\partial^{n_{i}}}{\partial
z_{i}^{n_{i}}}U^{r_{i}}(z_{i})U
^{-r_{i}}(z)U^{r}(z):  \nonumber \\ & = &
 :\prod_{i}U^{r_{i}(n_{i})}(z)U^{-r_{i}}(z)U^{r}(z):
\end{eqnarray}}

\begin{document}
\thispagestyle{empty}

\hfill DSF-T-93/44

$ \vspace{2cm} $

\begin{center}
{\LARGE\bf  Structure of Lorentzian algebras~}\\
{\LARGE\bf and}\\
{}~\\
{\LARGE\bf  Conformal Field Theory }
\end{center}

{}~~~~
\begin{center}
{\large\bf V. Marotta $\hspace{.5cm}$  A. Sciarrino } \\
{}~\\
\large
{ Dipartimento di Scienze Fisiche  \\
 Universit\'a di Napoli ``Federico II'' \\
and \\
INFN, Sezione di Napoli }

\end{center}
{}~~~~~

\begin{abstract}
\large
 The main properties of indefinite Kac-Moody and Borcherds algebras,
 considered in a unified way as Lorentzian algebras, are reviewed.
The connection with the conformal field theory of the vertex operator
construction is discussed. By the folding procedure a class of subalgebras
is obtained.
\end{abstract}

\vfill

\begin{tabbing}
\small\bf
Postal address: Mostra d'Oltremare Pad.19-I-80125 Napoli, Italy \\
E:mail: \=  Bitnet VMAROTTA (SCIARRINO)@NA.INFN.IT \\
        \>  Decnet AXPNA1::VMAROTTA (SCIARRINO)
\end{tabbing}

\normalsize\rm

\pagebreak

\section{ Introduction }

 This paper is devoted to present the mathematical aspects of Lorentzian
algebras with emphasis on the deep connections between their vertex operator
realization and the conformal properties of the fields which are introduced in
order to build up this representation.

Lorentzian algebras include the indefinite Kac-Moody algebras \cite{1} and the
generalized Kac-Moody algebras introduced by Borcherds \cite{23} which we call
Borcherds algebras.

For a review of the properties of Kac-Moody algebras (KMA) see
\cite{2}\cite{3}.

The history of Kac-Moody algebras development is linked to string theories and
conformal field theory, therefore the main applications are in these context
and make use of Affine KMA.

 However the interest in physics for the indefinite KMA has been emphasized by
many authors. In particular Julia \cite{4} has suggested that such algebras
may appear in the dimensional reduction of supergravity models and recently
Nicolai \cite{5} has shown that an hyperbolic extension of SL(2,R)
appears in the dimensional reduction of N = 1 supergravity from four to one
dimension.

The main features of this algebra, which we call $\hat{A}_{1}^{(1)}$, are
 recalled in Appendix A.

In string theories the Lorentzian algebras have appeared in different ways.
One of this was suggested by Goddard and Olive \cite{7} in the context of a
unification view of the heterotic string.

The algebras involved are very special kinds of Lorentzian algebras, one of
these is $ E_{19} $ the natural candidate for unification of
$ E(8)\otimes E(8) $ and $ SO(32) $ theories.

The use of Lorentzian algebras is necessary also in the fermionic string,
 see \cite{33} and  references therein, in which the role of ghosts sector
can be understood by means of these algebras.

Another way to introduce Lorentzian algebras in string theories consists in
reversing the sign of hantiolomorphic fields \cite{33}.
There is a strange interplay between this aspect and the role of ghosts in
previous application.

 The energy operator in physical models, exhibiting an Aff-KMA as symmetry
algebra, does not belong to the symmetry algebra but it can be identified with
the zero generator of the infinite dimensional Virasoro algebra associated
with the KMA. Slansky \cite{59} has recently suggested that the energy operator
as well as the particle number operator can be included in the symmetry algebra
 by an extension of the last one to generalized Kac-Moody algebra.

 Moreover the Monster Lie algebra \cite{30}  can be seen as an infinite rank
Borcherds algebra \cite{14}.

The strange coincidence that occurs between the theory of Monster and the
string theory is simply astonishing, so many physicists undergo the charm of
this fact.

We shall not discuss the Monster Lie algebra in this paper. For a discussion
of the connection of this algebra with the abstract theory of vertex algebra
and the conformal field theory see \cite{24}.

Nowadays most theoretical physicists are aware of the fact that the symmetries
of string field theory are larger then the symmetries of the related conformal
field theories (\cite{36} and references therein), but it is not well know
how this phenomenon can be realized.

We believe that all these, at present different and uncorrelated way to make
use of Lorentzian algebras in string theories, may be unified and understood
as a not  yet discovered symmetry of the quantized string field theory.

This paper is organized in the following way.
In sec. 2  we define a Lorentzian
algebra and we recall its main features. In sec. 3 we discuss a vertex operator
 realization of a Lorentzian algebra, mainly of the indefinite KMA, and we
emphasize the role of conformal symmetry.

In sec. 4 we discuss some class of not trivial subalgebras of the Lorentzian
algebra. At the end we point out some of the many yet open problems.

In Appendix A we recall in some detail the vertex operators of the hyperbolic
KMA ${\hat A}_{1}^{(1)}$ which we use as an example to illustrate general
features of the vertex realization of the Borcherds algebras as subalgebras.

In Appendix B we introduce an explicit system of simply roots for symmetric
hyperbolic KMA of rank $\geq 3 $.

\pagebreak

\section{ Lorentzian algebras}

\bigskip

We call Lorentzian algebra (LA) a Lie algebra whose roots belong to a lattice
in a $D$-dimensional space with indefinite metric
$ g_{\mu\nu} = (-, \dots, +, \ldots) $ with $q$ signs - and $p$ signs $ + $. In
the following we consider mainly the case $p=D-1$, $q=1$, but at the end we
briefly comment on the general case. There are two main classes of LA:
the indefinite KMA and the generalized KMA introduced by Borcherds which we
shall call in the following Borcherds algebra (BA).

  Firstly let us recall a few definitions which we need to define respectively
a rank $d$ KMA and BA \cite{1,23}. The rank $d$  is always equal to the
dimension $D$
of the space for Hyp-KMA (see below for definition) while in the general case
we have $ d \geq D$. In the case of $D = 26$ we are not considering the
infinite rank Monster Lie algebra \cite{30} although this seems the most
interesting case.

 A $ (d \times d) $ matrix $ A = [a_{ij}] $ is called a generalized
Cartan matrix
(GCM) \cite{3} if it satisfies the following conditions:

\begin{description}

\item[i)] \( a_{ij} \in   Z \)

\item[ii)] \( a_{ii}=2 \)

\item[iii)] \(  a_{ij} \leq 0 \:\:\:(i \neq j) \)

\item[iv)] \(   a_{ij} = 0 \:\:\: implies \:\:\: a_{ji} = 0 \)
\end{description}

 A matrix A is called  symmetrisable if the matrix UA is
symmetric, U being an invertible diagonal matrix. In the following we mainly
consider symmetric GCM, i.e simply laced Lie algebras.

A $ (d \times d) $ symmetric matrix
$ A = \mid a_{ij} \mid $ is called a generalized
symmetrized Cartan matrix (SCM) \cite{23} if it satisfies the above conditions
i), iii), iv) while condition ii) is replaced by

\begin{description}

\item[iia)] \( a_{ii} \in   Z \)

\item[iib)] \(   2a_{ij}/a_{ii} \:\:\: is \:\: integer \:\: if \:\: a_{ii}
\:\: is \:\:  positive \)

\end{description}

In the following we shall assume that

\begin{description}

\item[iic)] \( a_{ii} \:\:\: is \:\: 0 \:\: or \:\: 2 \)

\end{description}

A matrix A is called indecomposable if it cannot be reduced to a block
diagonal form by shuffling rows and columns.
To a GCM or to a SCM we associate a Dynkin diagram (DD), denoted sometimes
S(A), with the following properties:

\begin{description}

\item[a)]  S(A) has  $ d $  vertices

\item[b)] if $ a_{ij}a_{ji} = n \leq 4 $ , the vertices $ i $ and $ j $
 are joined by
$ \mid a_{ij} \mid \geq \mid a_{ji} \mid $ lines

\item[c)] if $ \mid a_{ij} \mid \geq \mid a_{ji} \mid $
($ \mid a_{ij} \mid \leq \mid a_{ji} \mid $), we put on the lines ($ ij $) an
arrow pointing, resp.,  towards the vertex $ j $ ($ i $)

\item[d)] if $ n  > 4 $ the vertices $ i $ and $ j $ are connected by a
boldface line on which
   an ordered pair of integers, $ \mid a_{ij} \mid $ and $ \mid a_{ji} \mid $,
is written.

\item[e)] if $ a_{ii} = 2 $  the i-th vertex will be denoted by a white
circle; if $ a_{ii} = 0 $ the i-th vertex will be denoted by a crossed
circle

\end{description}

Note that sometimes in the literature when condition b) is satisfied the
vertices are joined by $ n $ lines.

Clearly A is indecomposable if and only if the corresponding S(A) is a
connected diagram.

 To a given GCM A we associate a complex Lie algebra defined by $ 3d $
 generators,
$ E_{i}, \: F_{i} $ and $ H_{i} $ which satisfy the following
 commutation and Serre relations

\begin{eqnarray}
\left[ E_{i}, F_{j}\right] & = &  a_{ij} H_{i}  \label{eq: 6}  \\
\left[ H_{i}, H_{j}\right] & = &  0  \label{eq: 5}  \\
\left[ H_{i}, E_{j}\right] & = &  a_{ij} E_{j} \label{eq: 4} \\
\left[ H_{i}, F_{j}\right] & = & -a_{ij} F_{j}  \label{eq: 3} \\
(adE_{i})^{1-a_{ij}}E_{j} & = & (adF_{i})^{1-a_{ij}}F_{j} = 0 \:\:\:\:
 (i \neq j) \label{eq: 7}
\end{eqnarray}

To a given SCM A we associate a complex Lie algebra defined by $ 3d $
 generators,
$ E_{i}, \: F_{i} $ and $ H_{i} $ which satisfy the above
 commutation and Serre relations if $ a_{ii} \geq 0 $ and by the following
condition if $ a_{ii} = 0 $

\begin{equation}
 a_{ij} = 0 \rightarrow [E_i, E_j]  =  [F_i, F_j] = 0  \label {eq: BR}
\end{equation}

The algebra can be written
under the following form (triangular decomposition)

\[ G(A) = N_{-} \oplus  H  \oplus N_{+} \]

where H is the cartan subalgebra and $ N_{-} (N_{+}) $  are resp. the linear
span of $ F_{i} (E_{i}) $.

We have for LA detA $ \leq $ 0.
We recall that det A = 0 with rank of A equal to $ d-1 $ and determinant of any
leading principal submatrix positive, corresponds to affine KMA (Aff-KMA).

No general classification scheme exist for general Ind-KMA or BA.

The hyperbolic (Hyp) KMA algebras are a particular case of the Ind-KMA with
the further condition that every leading submatrix decomposes
into constituents of finite and/or affine type or, in equivalent
way, if deleting a vertex of the corresponding S(A) one gets DD of finite
or affine KM algebras.

  A classification of hyperbolic algebras has been made in refs. \cite{11,12}
generalizing previous results obtained in \cite{3}. In \cite{11} all
the DDs have been drawn, finding 238 DDs (of rank $ \geq 3 $) of which
142 DDs correspond to symmetric or symmetrisable GCM. The highest rank of the
Hyp-KMA is 10 and to this class belongs $ E_{10} $ which has appeared several
time in the context of string theories.
The determinant of a Hyp-KMA is always negative.

The root lattice $ \Gamma $ is generated by the elements $ \alpha_i $
($ i = 1,2,..,d$), called simple roots (SR), and  $ \Gamma $ has a real-valued
bilinear form defined by

\begin{equation}
a_{ij} = (\alpha_i, \alpha_j) = \alpha_{i} \cdot \alpha_{j}
\label{eq: 10}
\end{equation}

We have:

\newtheorem {PROOF}{Proposition}

\begin{PROOF}
 The determinant of a symmetrizable  Cartan
matrix A
corresponding to a Lie algebra with no imaginary SR of rank 3 is given by
\begin{equation}
 det|a_{ij}| = 2\{4 - (\alpha \beta \gamma)^{1/2} - (\alpha + \beta + \gamma)\}
\end{equation}
where $ \alpha, \beta, \gamma $ are, respectively, the number of lines joining
the 3 vertices of S(A).
\end{PROOF}

{\em Proof:}
 From the relation between S(A) and the matrix A we have that
$\alpha, \beta, \gamma$ are related to the off diagonal elements of the Cartan
matrix by $\alpha = a_{12}a_{21}$, $ \beta = a_{13}a_{31}$,
$\gamma = a_{23}a_{32}$. Defining the quantity
$\epsilon_{ij} = a_{ij}/a_{ji}$ which denotes the ratio between the lengths
of the i-th and j-th SRs and satisfies
$\epsilon_{ij} = \epsilon_{il} \epsilon_{lj}$ (no sum over $ l$) and
$\epsilon_{ij} = 1/\epsilon_{ji}$ we have
\begin{equation}
 a_{12}a_{23}a_{31} = \left (\frac{\alpha \beta \gamma}{\epsilon_{21}
\epsilon_{32}\epsilon_{13}}\right )^{1/2} = (\alpha \beta \gamma)^{1/2}
\end{equation}

\begin{equation}
 a_{13}a_{21}a_{32} = \left (\frac{\alpha \beta \gamma}{\epsilon_{31}
\epsilon_{12}\epsilon_{23}}\right )^{1/2} = (\alpha \beta \gamma)^{1/2}
\end{equation}
\noindent inserting the above equality in the expression of the determinant
of a rank 3 matrix we find the quoted result.
As the value of $det|a_{ij}|$ is an integer, in the case of loop DD we have
some constraints as $ (\alpha \beta \gamma)^{1/2} $ has to be an integer.
{\em Q.E.D.}

A root is called positive if it is a sum with not negative integers of SR.
A root $ \alpha $ is called $ {\sl real} $ if
$ (\alpha, \alpha) $ is positive , $ {\sl imaginary} $ if
$ (\alpha, \alpha) \leq 0 $.

Clearly the set of roots $\Delta $ satisfies

\begin{equation}
\Delta = \Delta ^{re} \cup \Delta ^{im}
\end{equation}

where

\begin{eqnarray}
\Delta ^{re} & = & \{ r\in \Delta \: : \:\: (\alpha,\alpha)=2 \:\: \} \\
\Delta ^{im} & = & \{ r\in \Delta \: : \:\: (\alpha,\alpha)\leq 0 \:\: \}
\end{eqnarray}

The multiplicity ($ mult $) of roots does not depend only from its length and
it can be computed by means of the Peterson recursion formula for LA whose
determinant of the defining GCM or SCM is negative:

\begin{equation}
(\alpha, \alpha-2\rho )C_{\alpha} = \sum_{\alpha'+ \alpha''=\alpha}
(\alpha', \alpha'')C_{\alpha'}C_{\alpha''}
\end{equation}
where
\begin{equation}
\alpha  \in \Gamma ^{+} ~~~~~~~~~~ \Gamma^{+}  =   \sum n_{i} \alpha_{i}
{}~~~~~~~~~~ n_{i} \in Z_{+}
\end{equation}
$ \rho $  is the ``unit" root
\begin{equation}
(\rho  ,\alpha_{i}) = 1/2 (\alpha_{i}, \alpha_{i})\:\:\:\:  \forall \: i \leq d
\end{equation}
and
\begin{equation}
C_{\alpha} = \sum_{n \geq 1} (1/n) mult(\alpha/n)
\end{equation}

The 23 DDs corresponding to symmetric Hyp-KMA of
rank $d \geq 3$ are reported in Appendix B where an explicit SR systems
is written.

Let us recall that an Aff-KMA can be obtained by adding to the lattice
$ \Gamma $ of roots of a finite Lie algebra $G_0$ (horizontal algebra) a
lightlike vector $ K^{+} $ :

\begin{equation}
 (K^{+},K^{+})=0 \:\:\:\:\: (K^{+},\alpha_{i})=0
\end{equation}

The affine root is obtained adding $ K^{+} $ to the lowest root of $G_0$
 (affinization procedure).

A class of  Hyp-KMA is obtained adding to the lattice of roots of Aff-KMA
an other lightlike root $ K^{-} $ such that:

\begin{equation}
(K^{-}, \alpha_{i}) = (K^{-}, K^{-}) = 0 \:\:\:\:    (K^{-}, K^{+}) = 1
\end{equation}

We call hyperaffinization or double-affinization the procedure of adding to
an Aff-KMA a SR
containing $ K^{-} $ such that this new root has scalar product equal to
$ -1 $
with the affine root and zero with the other SRs.

A classification of double-affinized KMAs has been obtained by Ogg \cite{19}.
In the original paper the procedure has been called superaffinization but we
prefer to change the name in hyper-affinization to avoid confusion with the
procedure of affinization of superalgebras.
One can show that the determinant of a  hyper-affine LA is just the
opposite of the determinant of the horizontal Lie algebra.
 Let us remark that in the case of Aff-KMA $ K^{+} $ is a root (not simple),
while in the case of Hyp KMAs $ K^{-} $ is not always a root.

  We recall that a graded Lie algebra $ G = \oplus^{\infty}_{-\infty}  G_{i}$,
generated
by $ G_{0} \oplus G_{1} \oplus G_{-1} $, simple (not containing non trivial
homogeneous ideals)
is said to be of finite growth \cite{13} if the dimension of the space
$ G_{i} $ grows as a power of $ \mid i \mid $. From Kac's theorem in \cite{13}
we see that the Hyp-KMA are of not finite growth.

Let us remark that the peculiar properties of the lightlike vectors $K^{\pm}$
allow one to build trans-hyperbolic KMA which contains a ``cluster" of Aff-KMA
which can be obtained taking the
direct sum of $n$ affine algebras $G_i$ (i=1,2,..,n) then by adding an extra SR
\begin{equation}
\alpha = -(K^+ + K^-)
\end{equation}
which will be simply connected with all the affine SR of any $G_i$.
The most celebrated example of such a ``cluster" is the trans-hyperbolic
algebra $E_{19}$ introduced by Goddard-Olive which in a $19$ rank Ind-KM which
contains as subalgebras $E_{8}^{(1)}\times E_{8}^{(1)}$ and $SO^{(1)}(32)$.
It has been argued the $E_{19}$ could be the underlying unifying symmetry
between the two consistent models of heterotic string. We remark that
in $10-dim$ Minkowskian space we can have a ``cluster" which is a rank 11
trans-hyperbolic KMA which contains as subalgebras
$SO(8)^{(1)}\times SO(8)^{(1)}$ and $SO^{(1)}(16)$.
In general in a  $D-dim$  one may build up cluster
of $p$ copies of the same Aff-KMA $G^{(1)}$ of rank $d$ ($pd = D-2$)
exhibiting a $p$-ality property. The determinant of the GCM A corresponding
to a ``cluster" is vanishing as one root is linearly dependent from the
others, in fact we have a LA with rank equal to $(D + 1)-dim$ space.
Finally we recall a few definitions and properties of the Weyl group for
KMA and BA.

We recall \cite{30}, that the Weyl group ($W$) of a KM algebra is a discrete
group of isometries of the dual of the Cartan subalgebra generated by the
reflections with respect to the SRs (fundamental reflections). For KMA The
elements
of W which are obtained as a product of an even number of fundamental
reflections form a normal subgroup  $W_0$ of $W$, called either conformal Weyl
group or even subgroup. Clearly, for Hyp KMAs,  $W_0 \in  SO^{+}(d-1,1) $,
while $ W \in  O^{+}(d-1,1) $.

In general the Weyl group is generated by the $d$ fundamental reflexions whose
action on the SRs is given by:

\begin{equation}
w_{\alpha_{i}}(\alpha_{j})= \alpha_{j}-a_{ij}\alpha_{i}
\end{equation}

Let us remind a few important properties of the Weyl group:

\begin{itemize}

\item the bilinear form $ ( \cdot \: , \cdot ) $, the sets $ \Delta $  and
$ \Delta ^{im} $ are invariant under the action of W;

\item the dimension of the space of a root is equal to the dimension of the
space of the reflected root;

\item contrary to what happens in the case of finite Lie algebras, in the root
  space of a KMA or of a BA may exist fixed points.

\end{itemize}

$\vspace{3cm}$

\section{ Vertex construction and Virasoro algebra }

 In this section we shall consider the relationship between  vertex
construction of Lorentzian algebras and conformal field theory (CFT)\cite{28}.

 Our motivation for doing this is to prove that the symmetries generated by
the Virasoro algebra that we associate to LA are necessary to obtain the
correct commutation relations. In fact we shall shown that a subalgebra of the
full
conformal algebra is relevant in the structure of the vertex construction of
LA.

 The understanding of the role of these symmetries is an important point in
physical applications of LA because the very strange character of these
algebras links many different and apparently uncorrelated topics in physics,
as we have briefly recalled in the Introduction, so it is tempting to suppose
that many yet unknown links exist between these topics.

 We consider in particular a bosonic CFT defined on a Lorentzian lattice and
then we construct the associate Virasoro algebra.

\subsection{ Virasoro algebra on Lorentzian lattice } \label {subsec_Conf}

 In the vertex construction of Lorentzian algebras we use a bosonic conformal
field theory on lattice, therefore in this section  we consider a generic
stress tensor $ T(z) $.

 Expanding in Laurent serie we obtain:

\begin{equation}
T(z)=\sum_{m}L_{m}z^{-m-2}
\end{equation}

where:

\begin{equation}
L_{m} = \frac{1}{2\pi i}\oint_{C_{0}}T(z)z^{m+1}dz = \frac{1}{2}\sum_{n}:
\alpha_{n}\cdot \alpha_{m-n}:
\end{equation}

with

\begin{equation}
\left[L_{m},L_{n}\right]=(m-n)L_{m+n}+\frac{D}{12}m(m^{2}-1)
\delta_{m+n}
\end{equation}

We consider also the subalgebra of projective transformations (Witt algebra):

\begin{equation}
\left[L_{m},L_{n}\right]=(m-n)L_{m+n}\hspace{2cm} \forall \:\: m,n\in \{\pm 1
,0 \}
\end{equation}

and the relative representations.

 These are the only globally invertible transformations on the Riemann sphere
\cite{28}.

 A primary field $\phi (z) $ (PF) satisfies the relations:

\begin{equation}
\left[L_{m},\phi (z)\right]=z^{m+1}\frac{d}{dz}\phi (z)+(m+1)h z^{m}
\phi (z) \:\:\:\:\forall\: m \label{eq: LF}
\end{equation}

then the relative coefficients of the Laurent expansion:

\begin{equation}
\phi (z)=\sum_{n} A_{n}z^{-n-h}
\end{equation}

have the following commutation relations with Virasoro algebra:

\begin{equation}
\left[L_{m},A_{n}\right]=\left[m(h-1)-n\right]A_{n+m} \:\:\:\:
\forall\: m,n \label{eq: 8}
\end{equation}

 A quasiprimary field (QPF) satisfies the same relations only with the
projective transformations subalgebra.

 The fields that does not satisfy the eq.(\ref {eq: LF}) and eq.(\ref {eq: 8})
relations are denoted as a secondary (SF).

 Let us define also a quasisecondary (QSF) field, provided it does not
satisfy the
relations of eq.(\ref {eq: LF}) restricted to $ m=\{ \pm 1 , 0\}$.

 Any QSF can be constituted by a set of coordinative derivates of QPF.

 Expanding this basic derivatives fields in Laurent serie we define:

\begin{equation}
i\frac{d^{l}}{dz^{l}}\phi (z)= \sum_{n} A_{n}^{(l)} z^{-n-l-h}
\end{equation}

 The $A_{n}^{(l)} $ coefficients are related to $ A_{n} $ by means of the
relations:

\begin{equation}
i\frac{d^{l}}{dz^{l}}\phi (z) = i\sum_{n} l!\left(\begin{array}{cc}-n-h\\l
\end{array}\right ) A_{n}z^{-n-l-h}
\end{equation}

\begin{equation}
 A_{n}= \frac{1}{2\pi i\: l!\left(\begin{array}{cc}-n-h\\l\end{array}\right)}
\oint_{C_{o}} dz z^{n+l+h-1} \frac{d^{l}}{dz^{l}}\phi (z)
\end{equation}

 Therefore we can extract the relations:

\begin{eqnarray}
 A_{n}^{(l)} & = & i l!\left(\begin{array}{cc}-n-h\\l\end{array}\right) A_{n}
\nonumber \\
A_{n}^{(l)} & = & 0 \:\:\:\: \forall \:\: n\: : \:\: -h-l+1 \leq n \leq -h
\end{eqnarray}

\begin{equation}
\left[L_{m},A_{n}^{(l)}\right]=\left[m(h-1)- n \right]
A_{n+m}^{(l)} \:\:\:\:\forall \:\: n  \in [-h-l+1,-h]
\end{equation}

 Then the $A_{0} $ operators relative to PFs with conformal dimension
$h=1 $ commute with full Virasoro algebra, instead, the $ A_{0} $
operators relative QPFs with $h=1 $ commute only with the $m=\{\pm 1, 0\}$
$L_{m} $ generators.

 Note that all the $A_{0}^{(l)} $ operators relative to QSF of dimension
$ h+l = 1 $ are vanish. These are the zero modes of quasinull fields (QNFs) and
 in the following we shall show its role in the vertex construction.

 In sec.(\ref{subsec_Vertex})  we construct the generators of a Lorentzian
algebra by means of the zero modes $A_{0} $
relative to fields of $ h = 1 $ but as we have showed that only QPFs give
contribution to $A_{0} $ operators, therefore, we can conclude that this
realization of Lorentzian algebras commute with the subalgebra of projective
transformations.

If the full algebra is built up by $A_{0} $ operators relative to PFs of
$ h =1 $ the full Virasoro algebra commute with it, therefore, in this case
 the vertex realization conserve the conformal properties.
We shall see that such a construction cannot be obtained, except in the case
of $ D = 26 $, for any Lorentzian algebra.

 Indeed, it is possible to prove this property in the case of 26 dimension by
means of the no-ghost theorem \cite{20}.

\subsection{ Vertex operator construction } \label {subsec_Vertex}

   The well-known vertex construction of an Aff KMA can be generalized
to the case of Ind KMAs along the lines of the covariant construction of
Goddard and Olive \cite{7,8}.

Note that this construction does not apply to BA in general as
eq.(\ref{eq: BR})  is not satisfied.
The construction can be carried on when the conditions of eq.(\ref{eq: BR}) are
never verified, e. g. when one imaginary SR is connected with all the other
real SRs. In sec.\ref{sec_Rep} we shall discuss in detail this point in a
specific case.

In this section we examine this construction in particular and introduce a
formalism in which the remarkable properties of Lorentzian algebras are
more evident.

We study a bosonic string in a compactificated space-time background in which
we consider only the holomorphic modes.
For an introduction to vertex algebra see \cite{24}.

The structure of the vertex representation depend only by the value of the
scalar products on the lattice, therefore we can generalize our consideration
to the algebras with root lattices in Minkowskian space $ R^{(q,p)} $.

We restrict the discussion to the $ q=1 $ case.

  Let us introduce $ D$ Fubini-Veneziano fields
($ \mu = 1,..,D $)

\1\

where

\begin{equation}
 [q^{\mu}, p^{\nu}] = ig^{\mu\nu} \:\:\:\:\:\:\:\:\: sign(g) = (-, +, +,.., +)
 \label {eq: 35} \end{equation}

\begin{equation}
 \left [ \alpha ^{\mu}_{n}, \alpha ^{\nu}_{m}\right ] = n \delta_{n+m,0}
g^{\mu\nu} \:\:\:\:\:\:\:\:\: \alpha^{\mu \dag}_{m} = \alpha^{\mu}_{-m}
\label {eq: 36}
 \end{equation}

The above fields are defined on a $ D $-dim Minkowskian torus and satisfy
periodical boundary conditions on the lattice $ \Gamma $.

 In our construction we consider only symmetric Cartan matrix, therefore, this
can be used directly as metric tensor

\[ a_{ij}=\alpha_{i}\cdot \alpha_{j} \]

 We introduce a stress tensor:
\begin{eqnarray}
T(z) & = & \frac{1}{2}\sum_{i}^{D}
:\alpha_{i}\cdot Q^{(1)}(z)\alpha^{i}\cdot Q^{(1)}(z): \nonumber \\  & = &
\frac{1}{2}:\alpha^{i}\cdot Q^{(1)}(z)a_{ij} \alpha^{j}\cdot Q^{(1)}(z):
\end{eqnarray}

where we introduce a set of  $ D $ independent roots $\{ \alpha_{i} \}$ and the
 dual basis $\{ \alpha^{i} \}$.

Note that in the case $ d > D $ we must consider only a subset of simple roots
that are associate to a submatrix of rank $ D $ of the Cartan matrix.

Let us define

\2\

and

\begin{equation}
r \cdot Q^{(n)}(z) = \frac{i \: d^{n}}{n!dz^{n}}r \cdot Q(z)
\end{equation}

where :  : denotes normal ordered product and the `polarization' $ r $ is in
$ \Gamma $. In general $r$ is an element of the lattice $\Gamma$, so a linear
combination of $\alpha_{i}$, but not necessarily a root. In the expression of
the VO of simple root $r$ denotes the SR $\alpha_{i}$.

$ U^{r}(z) $ are the vertex operator (VO) which are introduced for the vertex
realization of an Aff KMA.
It is possible introduce generalized vertex operator (GVO), see Borcherds
\cite{14} and \cite{20} by means of the following ordered product

\begin{equation}
:r_{1} \cdot Q^{(n_{1})}(z)r_{2} \cdot Q^{(n_{2})} \ldots r_{N} \cdot
Q^{(n_{N})}(z) U^{r}(z):
\end{equation}

where $ n_{i} \in Z_{+} $.
It is convenient to express a GVO in a different basis  \cite{15}
introducing a set of Schur polynomials, which are defined by the following
formal expansion

\begin{equation}
exp(\sum_{m>0} c_{m}z^{-m}) = \sum_{n}P_{n}(c) z^{-n}
\end{equation}

where $ c_{m} $ are commuting variables.
So we have

\begin{equation}
P_{n}(c) = \sum_{k_{l}} \frac{1}{k_{l}!}\left ( c_{l}\right )^{k_{l}}
\:\:\:\:\: \sum_{l}lk_{l} = n
\end{equation}

We use a new basis for the derivative fields in which they are represented as
Schur polynomials in the fields $ r \cdot Q^{(l)}(z) $ ($ 1 \leq l \leq n $):

\begin{eqnarray}
P_{n}(r\cdot Q^{(l)}(z)) & = & :\left(\frac{1}{n!}
\frac{d^{n}}{dz^{n}}U^{r}(z)\right)U^{-r}(z):  \nonumber \\
 & = & \lim_{z_{1}\rightarrow z}\frac{1}{n!}:\frac{\partial^{n}}{\partial z
_{1}^{n}}U^{r}(z_{1})U^{-r}(z):
\end{eqnarray}

The above equation has to be read as a Schur polynomial, the variable
$ c_{l} $ being now replaced by the field $ r \cdot Q^{(l)} $.
It follows that a GVO is an ordered product of Schur polynomials and standard
VO:

\begin{equation}
U^{\{r,r_{i}\}}_{\{(n_{i})\}}(z) = : \prod_{i} P_{n_{i}} (r_{i} \cdot Q^{(l_{i}
)}(z))U^{r}(z) : \label{eq: VO}
\end{equation}

which can be explicitly written in the following form:

\5\

 To any root of length L = $ r^{2} $ we associate a GVO with conformal weight
$ h = 1 $ such that

\begin{equation}
h = \frac{r^{2}}{2} + \sum_{i} n_{i} = 1  \label{eq: RT}
\end{equation}

So for L = 2 the corresponding GVO is the standard VO while for L = 0 it
is the photonic VO  \cite{7}.

Let us point out that the eq.(\ref{eq: RT}) is connected with the conformal
symmetry of the GVOs, moreover not all the set of $ (r_{i},n_{i}) $, which
satisfy eq.(\ref{eq: RT}), really appear in the construction of the algebra,
see below.

We can make a Laurent expansion of a GVO

\begin{equation}
U^{\{r, r_{i}\}}_{\{(n_{i})\}}(z) = \sum_{m} A^{\{r, r_{i}\}}_{m\{(n_{i})\}}
z^{-m-1}  \label{eq: LE}
\end{equation}

We select the terms in eq.(\ref{eq: LE}) with m = 0

\begin{equation}
A^{\{r,r_{1},...,r_{N}\}}_{\{(n_{i}),...,(n_{N})\}}=\frac{1}{2\pi i}\oint_{
C_{0}}dzU^{\{r,r_{1},...,r_{N}\}}_{\{(n_{1}),...,(n_{N})\}}(z)
\end{equation}

where the integral is performed along a closed path Co, including the
point $ z $ = 0.

We can summarize the relevant commutation relations in the following formula,
where suitable cocycles are supposed to have been included in the l.h.s. of
the equation

\begin{equation}
\left[A^{\{r,r_{i}\}}_{\{(n_{i})\}},A^{\{s,s_{j}\}}_{\{(n_{j})\}}\right]=
\frac{1}{(2\pi i)^{2}}\oint_{0}dz\oint_{z}d\xi U^{\{r,r_{i}\}}_{\{(n_{i})\}}
(z)U^{\{s,s_{j}\}}_{\{(n_{j})\}}(\xi)     \label{eq: AA}
\end{equation}

In particular if $ s = -r $ and $ s_{i} = -r_{i} $ we get

\begin{equation}
\left[A^{\{r,r_{1},...,r_{N}\}}_{\{(n_{1}),...(n_{N})\}},A^{\{-r,-r_{1},...,
-r_{N}\}}_{\{(n_{1}),...,(n_{N})\}}\right]=\chi^{\{r,-r,r_{1},-r_{1},...,r_{N},
-r_{N}\}}_{\{(n_{1}),(n_{1}),...(n_{N}),(n_{N})\}} r\cdot p \label{eq: EF}
\end{equation}

The $ \chi $ coefficients can be explicitly computed as a combination of
factorials

\begin{eqnarray}
\chi^{\{r,s,r_{i},s_{j}\}}_{\{(k_{i}),(k_{j})\}} & = & \prod^{N}_{i}
\prod^{M}_{j} \sum^{k_{i}}_{l_{i}=0} \sum^{k_{j}}_{l_{j}=0} (-1)^{l_{j}}
 \left ( \begin{array}{c} s_{j}\cdot (r-r_{i}) \\ k_{j} - l_{j} \end{array}
\right )
 \left ( \begin{array}{c}  r_{i}\cdot s_{j} \\ l_{i} \end{array} \right)
\nonumber \\
 & & \left ( \begin{array}{c}  r_{i}\cdot (s-s_{j}) \\ k_{i} - l_{i}
\end{array} \right )
 \left( \begin{array}{c}  r_{i}\cdot s_{j} - l_{i} \\ l_{j} \end{array} \right)
\end{eqnarray}

To compute the r.h.s. of the eq.(\ref{eq: AA})
 we have to evaluate the residues at poles of order

\begin{equation}
p = \sum k_{i}+\sum k_{j}-r\cdot s \:\:\:\:\:\:\:\: for \:\: z=\xi
\end{equation}

in the OPE.

{}From eq.(\ref{eq: RT}) for roots $ r $ and $ s $
it follows that the r.h.s. of eq.(\ref{eq: AA}) has poles

\begin{equation}
\frac{4 - (r + s)^{2}}{2} \geq p \geq - r\cdot s \geq 0
\end{equation}

therefore, there are no poles for

\begin{equation}
(r + s)^{2} \geq 4   \label{eq: NP}
\end{equation}

so we obtain a condition on the length of roots which insures the absence
of poles.

When eq.(\ref{eq: NP}) is satisfied the r.h.s. of eq.(\ref{eq: AA})
vanishes.

Let us remark that not all the operators are relevant for the construction
of the algebra as there is a class of GVOs which vanishes. In fact if
$ U^{\{r, r_{i}\}}_{\{(n_{i})\}}(z) $ can be written as a total $ z $
derivative
of a GVO, then $ A^{\{r, r_{i}\}}_{\{(n_{i})\}} $ = 0.

This property is a consequence of the quasi-conformal symmetry, in fact the
above fields are QNFs.

We can also show that the commutator of two GVOs that satisfy the eq.
(\ref{eq: RT}) is a linear combination of GVOs that satisfy the same relation.
In fact, in the r.h.s of eq.(\ref{eq: AA}) we extract the zero modes from a
linear combination of GVOs in which the conformal weight is:

\begin{equation}
h=\frac{(r+s)^{2}}{2} + \sum_{j} n_{j} + \sum_{i} n_{i} - r\cdot s -1
\end{equation}

therefore by means of eq.(\ref{eq: RT}):

\begin{equation}
h = h_{r} + h_{s} -1 = 1
\end{equation}

{}From eq.(\ref{eq: EF}) we see the Cartan generators $ H_{i} $ are given
by $ \alpha_{i} \cdot p $. Note that for $ n_{i} = 0 $, for $ r_{i} = 0 $ and
$ r^{2} = 2 $ the $ \chi $ coefficient in the r.h.s. is equal to 1.

Note that in the LAs the level $ K $ of the Aff-KM subalgebras is not a
c-number but simply a weight, therefore the vertex construction cannot depend
by the particular value of $ K $ so as it happen in the Fenkel-Kac-Siegal
construction.

This aspect is pointed out in \cite{21} where is showed that the auxiliary
parafermionic fields, that are necessary in the case of $ K\neq 1$, are
naturally obtained by an opportune factorization of the vertex operators.

\subsection {The structure of representations space }

 In this section we shall discuss the structure of CFT Fock space in relation
with the structure of the representations of Lorentzian algebras and the
conformal properties.

 The whole representations space is composed by the Fock space built up by the
creation operators $a_{n}^{\mu} $ with $ n<0 $ and the vector space of the
eigenstates of momentum operator with eigenvalue on Lorentzian lattice.

 This space can be decomposed in representations of Virasoro algebra
associated to the Lorentzian algebra.

A state is called primary (PS), quasiprimary (QPS), secondary (SS) and so on,
if
 it is created by the action of the corresponding field on the vacuum state
\begin{equation}
\mid \alpha > = \phi_{\alpha} (0)\mid 0 >
\end{equation}

A PS satisfies

\begin{eqnarray}
L_{n}\mid \alpha > & = &  0 \:\:\:\: \forall\: n > 0  \nonumber \\
L_{0}\mid \alpha > & = & h \mid \alpha >
\end{eqnarray}

while a QPS satisfies
\begin{equation}
L_{1}\mid \alpha > = 0 \:\:\:\: L_{0}\mid \alpha > = h \mid \alpha >
\end{equation}

A secondary state (SS) (or quasisecondary state (QSS)) that satisfies the same
PS (or QPS) relations is denoted as a null state (NS) (or quasinull state
(QNS)).

In sec. \ref {subsec_Conf} we have shown that the Lorentzian algebra is built
up by $A_{0} $ operators relative to QPFs of conformal dimension $ h = 1 $,
therefore we have
\begin{equation}
[ L_{m},A^{\{ r,r_{i}\}}_{\{(n_{i})\} }]=0 \hspace{2cm} \forall \:\: m\: \in
\{\pm 1,0 \} \label {eq: 14}
\end{equation}

{}From eq.(\ref {eq: 14}) it follows that the application of generator
$ A^{\{ r,r_{i}\}}_{\{(n_{i})\} } $ on a state of a representation of
Lorentzian algebra cannot change the conformal properties of the state
restricted to projective properties.

So we can associate a QPS to every operator $A^{\{ r,r_{i}\}}_{\{(n_{i})\} } $
by the relation:

\begin{equation}
U^{\{ r,r_{i}\}}_{\{(n_{i})\} }(0)\mid 0 > = \mid r,r_{i}(n_{i}) >
\end{equation}

with the properties:

\begin{eqnarray}
L_{0} \mid r,r_{i}(n_{i}) > & = & \left ( \frac{r^{2}}{2} + \sum_{i} n_{i}
\right ) \mid r,r_{i}(n_{i}) > =  \mid r,r_{i}(n_{i}) > \nonumber \\
L_{1}\mid r,r_{i}(n_{i}) > & = & 0
\end{eqnarray}

We denote this space as $ P^{(1)} $.

This space include also the QNS of dimension $ h+l = 1 $, but we have showed
that the relative operator $  $ is vanishing, so we must consider the quotient
space $ P^{(1)}/P^{(1)}_{N} $ where $ P^{(1)}_{N} $ indicate the space of null
states.

 Note that the QSFs associated to QNSs are fundamental to obtain a correct
commutation relations in fact the symmetric part of the OPE of the fields in
the r.h.s. of eq.(\ref {eq: AA}) is a QNF that correspond to a QNS so we can
have a commutator only on the quotient space.

The projective symmetry assure us that these states are orthogonal to all QPSs.

 In the case of 26 dimension is possible to extend the above considerations to
the whole Virasoro algebra by means of no-ghost theorem \cite{20}, so we must
consider the quotient $ P^{1}/P^{1}_{N} $ (where $ P^{1} $, $ P^{1}_{N} $ is
the space of PS and NS of dimension $ h=1 $).

We can introduce also a GVO representation of highest weight states
\begin{equation}
\mid \Lambda > = \sum_{i} m_{i} \mid \Lambda_{i} >
\end{equation}
 where $\Lambda_{i} $ are the fundamental weights.

These states are created by the tachionic vertex operator:
\begin{equation}
V^{\{\Lambda\}} = \frac{1}{2\pi i} \oint \frac{dz}{z} U^{\{\Lambda\}}(z)
\end{equation}

where $\Lambda $ belongs to the weight lattice $\Gamma^{*} $

The highest weight representations of Lorentzian algebras are built up by
lowering operators acting on the highest weight states (see \cite{3}). To any
highest weight $ \Lambda $ we associate the PS of conformal dimension
$ h = \Lambda^{2}/2 $, so it is also a highest weight state also for the
Virasoro algebra (and for the projective subalgebra).

This conformal weight is, generally, not an integer value.
This can be a problem if we consider the properties of the corresponding
conformal field theories, but we do not discuss this topic which is linked to
the fusion rules of vertex operator algebras.

Considerable effort is spent on this subject by  physicists and mathematicians
in particular for the study of the Moonshine module.

Moreover, in the physical applications, we must use CFTs that are unitary and
with local OPE properties so we must combine these theories in a opportune way.

A general state of a highest weight representation is obtained  by application
of lowering operators so we have:
\begin{equation}
\lambda = \Lambda - \sum_{i} r_{i}
\end{equation}

where $r_{i}$ are roots.

The corresponding state have the form
\begin{equation}
\mid \lambda,\lambda_{j} (n_{j}) > = \lim_{z \rightarrow 0}
U^{\{\lambda,\lambda_{j}\}}_{\{(n_{j})\}}(z)\mid 0 >
\end{equation}

where $ \lambda_{j} $ are polarizations

therefore the relative GVO is:
\begin{equation}
V^{\{\lambda,\lambda_{j}\}}_{\{(n_{j})\}} = \frac{1}{2\pi i} \oint \frac{dz}{z}
U^{\{\lambda,\lambda_{j}\}}_{\{(n_{j})\}}(z)
\end{equation}

The action of the Lorentzian algebras on the representation GVO is a
generalization of the commutation relation eq.(\ref{eq: AA})

\begin{equation}
\left[A^{\{r,r_{i}\}}_{\{(n_{i})\}},V^{\{\lambda,\lambda_{j}\}}_{\{(n_{j})\}}
\right]=
\frac{1}{(2\pi i)^{2}}\oint_{0}dz\oint_{z}\frac{d\xi}{\xi} U^{\{r,r_{i}\}}
_{\{(n_{i})\}}
(z)U^{\{\lambda,\lambda_{j}\}}_{\{(n_{j})\}}(\xi)     \label{eq: VA}
\end{equation}

where we use the fact that $\lambda $ is in the dual-lattice so we have an
integer scalar product with the root $ r $ and the cocycle relations are
extended also to the dual-lattice.

Then this operator creates states that transform under the representation of
highest weight $\Lambda $.

The conformal properties of  the $ V^{\{\lambda,\lambda_{j}\}}_{\{(n_{j})\}} $
are a consequence of general properties of the GVO and can be obtained
immediately if we recognize that this vertex is the therm $ - h $ in
the expansion of the vertex field $ U^{\{\lambda,\lambda_{j}\}}_{\{(n_{j})\}}
(z) $

\begin{equation}
U^{\{\lambda,\lambda_{j}\}}_{\{(n_{j})\}}(z) = \sum_{m} A^{\{\lambda,
\lambda_{j}\}}_{\{(n_{j})\}m}z^{-m-h} ~~~~~ m\in Z + h \label{eq: VE}
\end{equation}

where $ h $ is the conformal weight and $ V^{\{\lambda,\lambda_{j}\}}_{\{
(n_{j})\}} = A^{\{\lambda,\lambda_{j}\}}_{\{(n_{j})\}-h} $
so the commutation relations with the Virasoro algebra generators are
\begin{equation}
\left[L_{m},V^{\{\lambda,\lambda_{j}\}}_{\{(n_{j})\}}\right]=\left[(m+1)h
-m\right]A^{\{\lambda,\lambda_{j}\}}_{\{(n_{j})\}m-h}
\end{equation}

where we must consider only the value of $ m \in \{ 0,\pm 1\} $.

By the action of Lorentzian algebra on this GVO we deduce the following
relation
\begin{equation}
h = \frac{\lambda^{2}}{2} + \sum_{j} n_{j} = \frac{\Lambda^{2}}{2}
\end{equation}

in fact in the eq. (\ref{eq: VA}) we extract the  $ - h $ mode  from the OPE
of the fields with conformal weights
\begin{equation}
h_{r} = \frac{r^{2}}{2} + \sum_{i} n_{i} = 1 ~~~~~~
h_{\lambda} = \frac{\lambda^{2}}{2} + \sum_{j} n_{j}
\end{equation}

and the weight of the resulting field is
\begin{equation}
h_{r+\lambda} = h_{r} + h_{\lambda} - 1 = h_{\lambda}
\end{equation}

Therefore the conformal weight of the representations are invariant for the
algebra, but they are not invariant for the action of Virasoro algebra.

Note that $ h \leq 0 $ for the fundamental representations of the hyperbolic
algebras while the Virasoro algebra invariance needs the condition $ h = 1 $
so we cannot have conformal invariant representations in this case.

\bigskip

\section{Subalgebras of Ind-KMA} \label {sec_Rep}

There is no general classification scheme for subalgebras of Ind-KMA or BA. Of
course by deleting one or more vertex of a S(A) associated
to a LA one finds regular subalgebras and, in particular, all the finite and
affine subalgebras
of Hyp-KMA have been classified in \cite{12}. We want to show that in complete
analogy with the Fin- and Aff-KMA cases the procedure of ``folding", which
exploits the symmetry of the DD, can be applied to the case of LA and in the
particular case of Hyp-KMA gives singular subalgebras of the same type.
In fact we have

\newtheorem{PROF}{Proposition}
\begin{PROF}
The DD obtained by folding the DD of a Hyp-KMA corresponds to a Hyp-KMA.
\end{PROF}
{\em Proof:}
Let us recall that by folding of a Fin- or Aff-KMA an algebra of the
same type is obtained and that by deleting one or more dots of a Hyp-KMA a Fin-
or Aff-KMA is obtained. Let DD' be the Dynkin diagram obtained by folding
the DD of a Hyp-KMA (see below for a list), then it is evident that by deleting
one dot of DD' a diagram describing a Fin- or Aff-KMA is obtained.
{\em Q.E.D.}

Even if the Hyp-KMA have been completely classified, at our knowledge, there
is no generally accepted convention to identify them. We propose the
following convention: to identify a rank $d$ Hyp-KMA we delete the $d$-th
vertex
from the corresponding DD, so we get a Fin- or Aff-KMA which can be identified
by the standard Cartan notation, and then we add in square brackets the
number(s) identifying the vertex(s) which are connected with the $d$-th vertex
with an exponent(s) (omitted if equal to 1) denoting the number of lines in the
connection. We omit to write the label $d$ in the square bracket if the
$d$-th SR has the same length than the connected SR(s), otherwise we write
$d$ before (after) the number(s) identifying the other SR(s) if the $d$-th
is greater (smaller) than the connected SR.
This convention is obviously not unique as it depends on the labelling of the
SRs in the Hyp-, Fin- and Aff-KMA. In the case of hyperaffine KMA we identify
the algebra by putting a hat on the notation of the corresponding Aff-KMA.

We list below the singular subalgebras which are obtained by ``folding"
the DD of the symmetric Hyp-KMA (see Appendix B). We follow the labelling
of Appendix B and in the other cases we follow the convention of ref.\cite{2},
labelling however the affine root by the $d$ label rather than by the $0$
label. In the following list the first number denotes the DD
identifying the algebra in App. B and then we use the above convention

\begin{itemize}
\item 2) ~~ $A_1^{(1)}[(2)^2] \supset A_1^{2,4} $
\item 7) ~~ $A_2^{(1)}[2,3] \supset B_2[1,(32)^2]$
\item 9) ~~ $ \hat{A}^{(1)}_{3} \supset \hat{C}_3^{(1)}$
\item 10) ~~ $ A_3^{(1)}[1,3] \supset B_3[1,(43)^2]  \supset B_2[(2)^2]$
\item 10) ~~ $ A_3^{(1)}[1,3] \supset G_{2}[(32)^{3}] \supset  A_1^{2,3}$
\item 10) ~~ $ A_3^{(1)}[1,3] \supset C_3^{(1)}[(41)^{2}] \supset B_2[(2)^2]
\supset A_1^{3,2}$
\item 12) ~~ $ \hat{D}^{(1)}_{4} \supset B_4[3]  \supset A_3[(34)^3]$
\item 13) ~~ $ D^{(1)}_4[1] \supset B^{(1)}_{3}[2] \supset G^{(1)}_2[1]
         \supset A^{(2)}_2[2] \supset A_{1}^{1,5} $
\item 13) ~~ $  D^{(1)}_4[1] \supset B^{(1)}_{3}[2] \supset D^{(2)}_3[1]
         \supset G_2[(13)^{2}] $
\item 14) ~~ $ \hat{A}^{(1)}_{5} \supset \hat{C}_4^{(1)}$
\item 15) ~~ $ \hat{D}^{(1)}_{5} \supset B_5[3]$
\item 17) ~~ $ \hat{E}^{(1)}_{6} \supset B_5[5]$
\item 18) ~~ $ \hat{D}^{(1)}_{6} \supset B_6[3]$
\item 19) ~~ $ \hat{A}^{(1)}_{7} \supset \hat{C}_5^{(1)}$
\item 21) ~~ $ \hat{D}^{(1)}_{7} \supset B_7[3]$
\item 22) ~~ $ \hat{D}^{(1)}_{8} \supset B_8[3]$
\end{itemize}

where we introduce a notation to individuate the Hyp-KMAs of rank 2 in which we
write the absolute values of the GCM elements $a_{12}$ and $a_{21}$ on the
symbol $A_1$.

The procedure of folding can also be applied to transhyperbolic KMA, e.g.
by folding $E_{19}$ we $E^{(1)}_8[(10,9)^2]$.

The folding can also be applied to BA and may give algebras of the same type
with associated Cartan matrix which are no more symmetric.

To give an example let us consider the following BA associated with the SCM

\[ A= \left( \begin{array}{crcrcrcrc}
\:\:\: 0 & -1 & \:\:\: 0 \\ -1 & \:\: 2 & -1 \\ \:\:\: 0 & -1 & \:\:\: 0
\end{array}\right) \]

the subalgebra generated by
\begin{eqnarray}
E_{\hat{\alpha}_{1}} & = & E_{\alpha_{1}} + E_{\alpha_{3}} \\
E_{\hat{\alpha}_{2}} & = & E_{\alpha_{2}} \\
\hat{H}_{1} & = & H_{1} + H_{3} \\
\hat{H}_{2} & = & H_{2}
\end{eqnarray}
which can be obtained by folding the roots $\alpha_{1}$ and  $\alpha_{3}$,
satisfies all the defining eqs. (\ref{eq: 6})-(\ref{eq: BR}) with the following
Cartan matrix

\[ A= \left( \begin{array}{crcr}
\:\:\: 0 & -2 \\  -1 & \:\: 2
\end{array}\right) \]

Note that one could have chosen also the second root as a vanishing one.

It is interesting to remark that the procedure of folding does not require the
two roots $\alpha_{1}$  and  $\alpha_{2}$ to be of the same type, i.e. both
imaginary or real.

In the example considered with $\alpha_{3}$ real by folding we get  $ B_{2} $.

Finally we remark that some BA may appear as subalgebra of Ind-KMA.
In fact the set of root of BA may be included in the set of roots
of a Ind-KMA, an essential difference being the fact that imaginary
roots appear as simple roots in BA while they appear at some height in
the Ind-KMA. We are not able to make a general discussion of this topic and
we discuss this point in some detail in a specific
example, namely the case of the Hyp-KMA $ \hat{A}^{(1)}_{1} $ of which
we recall in Appendix A the main properties and the VO realization.
In this specific case we illustrate also the difficulties which appear in the
construction of a VO for general BA.

Let us consider the Borcherds algebra, that we denote as $A_1^{(0)} $,
associated to the $ 2\times 2 $ SCM

\[ A= \left( \begin{array}{crcr}
\:\:\: 0 & -1 \\  -1 & \:\: 2
\end{array}\right) \]

and the corresponding S(A) is:
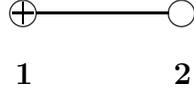
\begin{center}
{\normalsize\bf
\begin{picture}(200,200)
\bf
\put(50,100){\circle{10}}
\put(110,100){\circle{10}}
\put(55,100){\line(1,0){50}}
\put(46,100){\line(1,0){8}}
\put(50,96){\line(0,1){8}}
\put(47,73){1}
\put(107,73){2}
\end{picture}}
\end{center}
\normalsize\rm
The SRs are $ \hat{\alpha}_{1}, \hat{\alpha}_{2} $ which can be expressed
as linear combination of the SRs of $\hat{A}_1^{(1)}$, denoted by $ \alpha_i $,
see App. A
\begin{eqnarray}
\hat{\alpha}_{1} & = & \alpha_1 + \alpha_2 = K^+ \equiv [1, 1, 0]^0
\nonumber  \\
\hat{\alpha}_{2} & = & \alpha_3 = -(K^{+} + K^{-}) \equiv [0, 0, 1]^2
\label{eq: AP} \\
\end{eqnarray}

{}From the identification of the BA $A^{(0)}_1$ as a subalgebra of an
Hyp-KMA we can get a VO realization of the algebra and we can discuss
the structure of the fundamental representations.
In order to achieve this we have to show that with this choice we really build
up a BA, i.e. we have to
verify that the defining eqs.(\ref{eq: 6})-(\ref{eq: BR}) are satisfied when
we identify $ E_i $ with $ A^{\hat{\alpha}_{i}} $ which is given by the
eq.(\ref{eq: 95}) that in this case read

\begin{equation}
E_1 = A_1^{(\alpha_{1}+\alpha_{2},\alpha_{1}-\alpha_{2})} =
\frac{1}{2\pi i} \oint_{C_{0}}dz : \left(\frac{d}{dz} - \frac{d}{d\xi}
\right)U^{\alpha_{1}}(z)U^{\alpha_{2}}(\xi) : \mid_{z = \xi}
\end{equation}

\begin{equation}
E_2 = A^{\alpha_{3}} =
\frac{1}{2\pi i} \oint_{C_{0}}dz U^{\alpha_{2}}(z)
\end{equation}

Eq.(\ref{eq: 6}) ($ i \neq j$) is immediately verified just remarking that
$ (\alpha_i, -\alpha_j) = 1 $. Eq.(\ref{eq: 5}) ($ i = j$) is easily verified
just remarking that we have

\begin{equation}
H_i = \hat{\alpha_{i}} \cdot p = \hat{\alpha_{i\mu}}p^{\mu}
\end{equation}

and $ [p^{\mu}, \, p^{\nu}] = 0 $. Eqs.(\ref{eq: 4})-(\ref{eq: 3}) follow from
the property

\begin{equation}
{}~[A, \, B] = c \rightarrow [A, \, \exp{B}] =c \exp{B} ~~~~~~~~~ c \in {\bf C}
\end{equation}

and from eqs.(\ref{eq: 90}) and (\ref{eq: 91}). Eqs.(\ref {eq: 7}) (the Serre
relations) can be verified in a straightforward way and have been explicitly
computed in \cite{46}.
Let us remark that the fact that the elements of $ \hat{A}_1^{(1)} $ above
specified behave just as the generators of a BA can be shown quite in general
without any use of the VO realization.

A few words on the representations of this BA.
The fundamental weights are:
\begin{eqnarray}
\Lambda_1 & = & n \alpha - K^{+} + K^{-} = (1, 0) \:\:\:\: (n \in Z_{+})
\nonumber \\
\Lambda_2 & = & n \alpha - K^{+} = (0, 1)
\end{eqnarray}
With a suitable choice of $n$ ($n = 0$) we can find the fundamental
representations of $ A_1^{(0)} $ as subrepresentations of the fundamental
representations $(0, 1, 0)$ and $(0, 0, 1)$ of
$ \hat{A}_1^{(1)} $ discussed in \cite{46}.

\begin{equation}
(1, 0) \subset (0, 1, 0)  ~~~~~~~~~~~~  (0, 1) \subset (0, 0, 1)
\end{equation}

A few states of the (1, 0) representation are given in fig.1.
We can easily interpret the structure of the representation. The generator
$ E^{-\hat{\alpha}_1} $ corresponds to the {\sl grade changing} operator,
while the $ E^{-\hat{\alpha}_2} $ is the lowering generator of $ A_1 $. The
operator $ N $ associated to the commutator of the elements $ E^{+r}, E^{-r} $
($r = -2 \hat{\alpha}_1 - \hat{\alpha}_2 = - K^{+} + K^{-}$) is the
intertwining operator which connects different $ A_1 $ representations. The
multiplicity of the weights can be computed by the Freudental recursion formula
and it has been explicitly computed for the lowest multiplicity by Slansky
\cite{59}.

Let us remark  that the choice we have done for the SRs is not unique.
Indeed we could have made another choice for the 2nd SR, e.g.
$ \alpha_{2} + \alpha_{3} $ instead of $ \alpha_{2} $. Therefore
$\hat{A}_1^{(1)}$ contains many BAs of the same type as subalgebras. Moreover
one can also realize the the BA $ A_1^{(0)} $ in a form which is not subalgebra
of $\hat{A}_1^{(1)} $. In particular the structure of this BA which is
discussed
in \cite{59} and which is obtained by adding to the algebra $ A_1 $ a suitable
null root can be specified by the SRs:
\begin{eqnarray}
\hat{\alpha}_{1} & = & \alpha \nonumber \\
\hat{\alpha}_{2} & = & 1/2(-\alpha + K^{+} - K^{-})
\end{eqnarray}
These roots {\bf do not belong} to the set $ \Delta^{+} $ of the roots of
$\hat{A}_1^{(1)}$. A VO realization of this algebra should be of the form

\begin{equation}
V^{\{\hat{\alpha}_{2}\}}(z) = r \cdot Q^{(1)}(z) U^{\{\hat{\alpha}_{2}\}}(z)
\end{equation}

The polarization vector $ r = n\alpha + n_{+}K^{(+)} + n_{-}K^{(-)} $
is undetermined. By conformal properties we may require
\begin{equation}
r \cdot \alpha_2 = 0
\end{equation}
However this choice still leaves some arbitrariness for $ r $. The $ N $
operator should be identified with an element of the Cartan subalgebra,
but there is no affine subalgebra contained in this realization, as
$ K^{+} $ does not belong to the set of roots.

In order to emphasize that, even if the set of roots of BA is contained in the
set of an Ind-KMA, the BA not always can be considered as a subalgebra, let us
consider the BA, that we denote as $ A_2^{(0)} $ associated to the
$3\times 3$ SCM

\[ A= \left( \begin{array}{crcrcrcrc}
\:\:\: 0 & -1 & \:\:\: 0 \\ -1 & \:\: 2 & -1 \\ \:\:\: 0 & -1 & \:\:\: 2
\end{array}\right) \]
and the corresponding S(A) is:
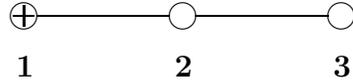
\begin{center}
{\normalsize\bf
\begin{picture}(300,200)
\bf
\put(50,100){\circle{10}}
\put(110,100){\circle{10}}
\put(170,100){\circle{10}}
\put(55,100){\line(1,0){50}}
\put(115,100){\line(1,0){50}}
\put(46,100){\line(1,0){8}}
\put(50,96){\line(0,1){8}}
\put(47,77){1}
\put(107,77){2}
\put(167,77){3}
\end{picture}}
\end{center}
\normalsize\rm
The simple roots can be chosen as:
\begin{eqnarray}
\hat{\alpha}_{1} & = & \alpha_1 + \alpha_2 = K^+ \equiv [1, 1, 0]^0
\nonumber  \\
\hat{\alpha}_{2} & = & \alpha_3 = -(K^{+} + K^{-}) \equiv [0, 0, 1]^2
\label{eq: AD} \\
\hat{\alpha}_{3} & = & \alpha_2 = - \alpha + K^{+} \equiv [0, 1, 0]^2 \nonumber
\end{eqnarray}
The BA can be build up adding to the (real) SRs of the Lie algebra $A_2$
an imaginary (null) root.
The VOs describing the elements of $\hat{A}_1^{(1)}$ corresponding to the
above (not simple) roots {\bf do not satisfy} the defining
eqs.(\ref{eq: 6})-(\ref{eq: BR})
of a BA so they do not form a realization of $A_2^{(0)}$. In fact, e.g.,
we have
\begin{equation}
{}~[E^{\hat{\alpha}_1},\, E^{\hat{\alpha}_3}] \neq 0 ~~~~~~~~~~~~ a_{13} = 0
\end{equation}

\begin{equation}
{}~[E^{\hat{\alpha}_1},\, E^{-\hat{\alpha}_3}] \neq 0
\end{equation}

Indeed as $E^{\hat{\alpha}_i}$ are described by GVOs, their commutator is
not vanishing even if the scalar product of the corresponding roots is
vanishing, as explicitly remarked in the case of $\hat{A}_1^{(1)}$.
This difficulty is overcome in the case of $A^{(0)}_1$ as there is no
SR unconnected with the imaginary SR.

Of course BA may contain as subalgebra KMA. We do not discuss
this point as, missing a VO or other type of representation of a BA, it is
not really useful.

\pagebreak

\section{Conclusions}

We have presented an essential review of the Lorentzian algebras with emphasis
on the connection between the conformal field theory language and the
properties
of the fields which appear in the VO.
The discussion of the conformal behaviour of the field requires
 the construction of the Virasoro algebra associated with the KMA, which
has been discussed, in full generality, by Borcherds in \cite{14}.
   The conformal structure of the fields in the vertex operator construction
is extremely relevant for physical applications and it depends on the value
of the dimension $ d $. An other relevant aspect, which is related with the
previous one, is the action of the GVOs on the Fock space of
 the representation.
Many problems are still opened. One fundamental point still missing is
the relation between a current algebra realization and the Lorentzian
algebras. Moreover while the interpretation of the VO corresponding
to imaginary negative roots as corresponding to massive states in the
string theory language is clear, one may wonder if
there is a deeper and more essential connection to be discovered
between Lorentzian algebras and CFT. In sec. 1 we recalled that one of
the motivation for the study of Lorentzian algebra has been based on
the need in the case of superstring to introduce fermionic ghosts in
order to insure the correct conformal behaviour and then to extend the
space of the lattice to a Lorentzian one. It is well known that ghosts
are essential ingredients in the quantization procedure. So it is tempting
to think that Lorentzian algebras are more closely connected with the
quantization procedure. Many speculations in this direction have appeared,
but there is no clear convincing argument for this conjecture. Very
likely this point is not completely unrelated from the previous one.

The bosonic VO construction we have presented in sec. 3 is completely
general and it applies to any Ind-KMA. In this paper we have discussed the case
of $D - dim $ space with indefinite metric $g_{\mu,\nu} = ( -, + \dots + )$.
The generalization to the case with more then one $(-)$ sign can be done
without difficulty just by a suitable redefinition of eqs.(\ref{eq: 35}) and
(\ref{eq: 36})

It is natural to argue that also
for this algebras a fermionic VO construction can be obtained which can
be essential for the case of not simply-laced algebras and, may be, to
get general VO representations of the BA.

The procedure of folding has allowed us to find a class of subalgebras and, in
analogy with the case of  Aff-KMA \cite{19}, may give a method to built VO
realization of not symmetric LA. A difficulty which has to be overcome is
connected to the construction of suitable auxiliary fields to obtain the
fermionic nature of the fields connected to the short roots and of the
necessary cocycles.

     The general structure of an IR for any Hyp KMA can be inferred from our
discussion in sec. 4. However many open problems are still present. For
instance a general proof of the complete reducibility of a h.w. IR in terms of
IRs of the affine subalgebra and a formula (at least formal) giving the
decomposition of an IR of Hyp KMA with respect to the affine subalgebra are
missing. Moreover the string functions, which in the case of Aff KMAs
allow the computation of the multiplicity of the weights, are not known for
the Hyp KMAs.

$ \vspace{.5cm} $

ACKNOWLEDGMENTS. We thank M. Durante and M.R. Nappo for the hospitality offered
and the valuable help give to one of us (VM) in computer drawing of the
Figures.

\pagebreak

\appendix{Appendix A}

 The symmetric GCM defining the $ \hat{A}^{(1)}_{1} $ algebra is \cite{3,6}:

\[ A= \left( \begin{array}{crcrcrcrc}
\:\:\: 2 & -2 & \:\:\: 0 \\ -2 & \:\: 2 & -1 \\ \:\:\: 0 & -1 & \:\:\: 2
\end{array}\right) \]

and the corresponding S(A) is:

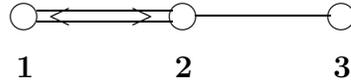
\begin{center}
{\normalsize\bf
\begin{picture}(300,200)
\bf
\put(50,100){\circle{10}}
\put(110,100){\circle{10}}
\put(170,100){\circle{10}}
\put(55,102){\line(1,0){51}}
\put(55,98){\line(1,0){51}}
\put(115,100){\line(1,0){51}}
\put(59,97){$<$}
\put(90,97){$>$}
\put(47,77){1}
\put(107,77){2}
\put(167,77){3}
\end{picture}}
\end{center}
\normalsize\rm

The simple roots are, denoting by $ \alpha $   the root of $ A_{1} $,
$ (\alpha_{i} , \alpha_{i}) = 2 $:

\begin{eqnarray}
\alpha_{1} & = & \alpha  \nonumber  \\
\alpha_{2} & = & - \alpha + K^{+} \label{eq: AS} \\
\alpha_{3} & = & - K^{+} - K^{-} \nonumber
\end{eqnarray}

 The algebra is defined by eq.(\ref{eq: 6}$ \div $\ref{eq: 7}), where now
$ E_{i}(F_{i}) $ (i=1,2,3) correspond resp. to $ \alpha_{i}(-\alpha_{i}) $
and  $ a_{ij} $
is given in terms of the SR by eq.(\ref{eq: 10}).

The set of roots $ r $ ( $ \Delta = \{ r \} $ ) \cite{6} is given by
$ r= \sum_{i=1}^{3} k_{i} \alpha_{i} $
where the triple of integers is constrained by:

\begin{equation}
(k_{1}^{2}+k_{2}^{2}+k_{3}^{2})-2k_{1}k_{2}-k_{2}k_{3} \leq 1
\end{equation}

In the following we will specify a root by a triple of integers
$ [k_{1},k_{2},k_{3}] $, denoting by a superscript on the triple its length.
 In \cite{46} a formula for computing the triple of (finite) numbers
$ [k_{1}, k_{2}, k_{3}] $ in function of the height ($ ht $)
$ (ht = k_{1}+k_{2}+k_{3}) $ and of the length (L) of
the roots is given.

The roots of low height ($ ht $) are:

\begin{description}

\item[ht = 1] $ \{ [1,0,0]^{2}; [0,1,0]^{2}; [0,0,1]^{2} \} $

\item[ht = 2] $ \{ [1,1,0]^{0}; [0,1,1]^{2} \} $

\item[ht = 3] $ \{ [2,1,0]^{2}; [1,2,0]^{2}; [0,2,1]^{0} \} $

\item[ht = 4] $ \{ [2,2,0]^{0}; [2,1,1]^{2}; [1,2,1]^{0} \} $

\item[ht = 5] $ \{ [3,2,0]^{2}; [2,3,0]^{2}; [2,2,1]^{-2}; [1,2,2]^{2} \} $

\end{description}

Remark the first vanishing (negative) length root does appear, respectively
at $ ht $ = 2, 5.

 Let us explicitly write a few relevant commutation relations for the simple
roots of $ \hat{A}^{(1)}_{1} $.

We have (i=1,2,3)

\begin{equation}
\left [ A^{\alpha_{i}}, A^{-\alpha_{i}}\right ] = \alpha_{i}\cdot p
\label {eq: 90}
\end{equation}

with

\begin{equation}
\left [ \alpha_{i}\cdot p , A^{\pm \alpha_{j}}\right ] =
\pm (\alpha_{i}\cdot\alpha_{j})A^{\pm \alpha_{j}} = \pm a_{ij}A^{\pm
\alpha_{j}}
\label {eq: 91}
\end{equation}

 where are been made use of eqs.(42), (33) and (31).

Then we have:

\begin{equation}
\left [ A^{\alpha_{1}}, A^{\alpha_{3}}\right ] = 0 \hspace{2cm} (a_{13}=0)
\end{equation}

\begin{equation}
\left [ A^{\alpha_{2}}, A^{\alpha_{3}}\right ] = A^{\alpha_{2}+\alpha_{3}}
\hspace{1cm} (a_{23}=-1)
\end{equation}

\begin{equation}
\left [ A^{\alpha_{2}} , \left [ A^{\alpha_{2}}, A^{\alpha_{3}}\right ]\right ]
= 0
\end{equation}

as ($\alpha_{2}\cdot(\alpha_{2}+\alpha_{3})=1 $)

\begin{eqnarray}
 & & \left[A^{\alpha_{1}},A^{\alpha_{2}}\right] =   \nonumber \\
 & = & \frac{\chi^{\{\alpha_{1},\alpha_{2}\}}}{2\pi i}\oint_{C_{0}}dz
:\frac{1}{2}\left(\frac{d}{dz}-\frac{d}{d\xi}\right)U^{\alpha_{1}}(z)U^{
\alpha{2}}(\xi):\mid_{z=\xi}
 \nonumber  \\  & = &
\frac{1}{2}A^{\{\alpha_{1}+\alpha_{2},\alpha_{1}-\alpha_{2}\}}_{(1)} ~~~~~~
(a_{12}=-2)  \label{eq: 95}
\end{eqnarray}

\begin{eqnarray}
\left [ A^{\alpha_{1}} , \left [ A^{\alpha_{1}}, A^{\alpha_{2}}\right ]\right ]
 & = & \left [ A^{\alpha_{1}} , \frac{1}{2}A^{\{\alpha_{1}+\alpha_{2},
\alpha_{1}-\alpha_{2}\}}_{(1)} \right ] =   \nonumber  \\
 & = & \frac{1}{2} (\alpha_{1}\cdot(\alpha_{2}-\alpha_{1}))
A^{2\alpha_{1} + \alpha_{2}}
\end{eqnarray}

 Let us remark that the above commutator is not vanishing  in spite of the
vanishing of the scalar product $\alpha_{1}\cdot(\alpha_{1}+\alpha_{2})=0 $
as the VO corresponding to the root $\alpha_{1}+\alpha_{2} $ is a GVO and
then in the  commutator a pole of order $\alpha_{1}\cdot(\alpha_{1}
+\alpha_{2})-1=-1 $ appears, see eq.(44). The commutator of $A^{\alpha_{1}}$
with $A^{2\alpha_{1} + \alpha_{2}}$ vanishes as
$\alpha_{1}\cdot( 2 \alpha_{1}+\alpha_{2}) = 2 $.

\pagebreak

\appendix{Appendix B}

We write a system of simple roots for all symmetric hyperbolic algebras
($ d\geq 3 $)
in which we decompose the Lorentzian lattice in a transverse Euclidean
 and a longitudinal-timelike lattice.

For any algebra we report the Dynkin diagram and we write the simple roots in
terms of the simple roots of the related affine subalgebra while the extended
roots is written in terms
of two light like vectors $ K^{\pm} $ in the longitudinal-time like space.
  In the following we denote with HR the highest root of the horizontal finite
Lie algebra

\begin{enumerate}

\item $ \hat{A}_{1}^{(1)} $

$ r_{i} \in A_{1}^{(1)}~~~ for ~ r_{i}\in \{1,....,2\}; $
$ ~~ r_{3}  =  - K^{+} - K^{-} $

\item $ A_{1}^{(1)}[(2)^{2}] $

$ r_{i} \in A_{1}^{(1)}~~~ for ~ r_{i}\in \{1,....,2\}; $
$ ~~ r_{3} = -\frac{1}{2} K^{+} - 2 K^{-} $

\item $  A_{1}^{(1)}[1,(2)^{2}]$

$ r_{i} \in  A_{1}^{(1)}~~~ for ~ r_{i}\in \{1,....,2\}; $
$ ~~ r_{3} = - \lambda_{s} - \frac{1}{2}K^{+} - 3 K^{-} $

\item $A_{1}^{(1)}[1,2]$

$ r_{i}  \in  A_{1}^{(1)}~~~ for ~ r_{i}\in  \{1,....,2\}; $
$ ~~ r_{3} = - \lambda_{s} - \frac{3}{8}K^{+} - 2 K^{-} $

\item $A_{1}^{(1)}[(1)^{2},(2)^{2}]$

$ r_{i} \in A_{1}^{(1)}~~~ for ~ r_{i}\in  \{1,....,2\}; $
$ ~~ r_{3} =  - HR - 4 K^{-} $

\item $  \hat{A}_{2}^{(1)}$

$ r_{i} \in A_{2}^{(1)}~~~ for ~ r_{i}\in  \{1,....,3\}; $
$ ~~ r_{4} = - K^{+} - K^{-} $

\item $A_{2}^{(1)}[2,3]$

 $ r_{i}  \in  A_{2}^{(1)}~~~ for ~ r_{i}\in \{1,....,3\}; $
$ ~~ r_{4} = - HR - 2 K^{-} $

\item  $ A_{2}^{(1)}[1,2,3]$

$ r_{i}  \in  A_{2}^{(1)}~~~ for ~ r_{i}\in  \{1,....,3\}; $
$ ~~ r_{4} =  - HR - 3 K^{-} $

\item $ \hat{A}_{3}^{(1)}$

$ r_{i}  \in  A_{3}^{(1)}~~~ for ~ r_{i}\in  \{1,....,4\}; $
$ ~~ r_{5} =  - K^{+} - K^{-} $

\item $ A_{3}^{(1)}[1,3]$

$ r_{i}  \in  A_{3}^{(1)}~~~ for ~ r_{i}\in  \{1,....,4\}; $
$ ~~ r_{5}  =  - HR - 2 K^{-} $

\item $  \hat{A}_{4}^{(1)}$

$ r_{i}   \in  A_{4}^{(1)}~~~ for ~ r_{i}\in  \{1,....,5\}; $
$ ~~ r_{6} =  - K^{+} - K^{-} $

\item $  \hat{D}_{4}^{(1)} $

$ r_{i}  \in  D_{4}^{(1)}~~~ for ~ r_{i}\in  \{1,....,5\}; $
$ ~~ r_{6} =  - K^{+} - K^{-} $

\item $  D_{4}^{(1)}[1] $

 $ r_{i} \in  D_{4}^{(1)}~~~ for ~ r_{i}\in  \{1,....,5\}; $
$ ~~ r_{6} =  - HR - 2 K^{-} $

\item $ \hat{A}_{5}^{(1)} $

$ r_{i}  \in  A_{5}^{(1)}~~~ for ~ r_{i}\in  \{1,....,6\}; $
$ ~~ r_{7}  = - K^{+} - K^{-} $

\item $ \hat{D}_{5}^{(1)}$

$ r_{i} \in D_{5}^{(1)}~~~ for ~ r_{i}\in  \{1,....,6\}; $
$ ~~ r_{7} = - K^{+} - K^{-} $

\item $ \hat{A}_{6}^{(1)}$

$ r_{i}  \in  A_{6}^{(1)}~~~ for ~ r_{i}\in  \{1,....,7\}; $
$ ~~ r_{8} =  - K^{+} - K^{-} $

\item $ \hat{E}_{6}^{(1)}$

$ r_{i}  \in  E_{6}^{(1)}~~~ for ~ r_{i}\in  \{1,....,7\}; $
$ ~~ r_{8} = - K^{+} - K^{-} $

\item $\hat{D}_{6}^{(1)}$

$ r_{i}  \in  D_{6}^{(1)}~~~ for ~ r_{i}\in  \{1,....,7\}; $
$ ~~ r_{8}  =  - K^{+} - K^{-} $

\item $ \hat{A}_{7}^{(1)}$

$ r_{i}  \in  A_{7}^{(1)}~~~ for ~ r_{i}\in  \{1,....,8\}; $
$ ~~ r_{9}  =  - K^{+} - K^{-} $

\item $\hat{E}_{7}^{(1)}$

$ r_{i}  \in  E_{7}^{(1)}~~~ for ~ r_{i}\in  \{1,....,8\}; $
$ ~~ r_{9} =  - K^{+} - K^{-} $

\item $ \hat{D}_{7}^{(1)}$

$ r_{i}  \in  D_{7}^{(1)}~~~ for ~ r_{i}\in  \{1,....,8\}; $
$ ~~ r_{9}  =  - K^{+} - K^{-} $

\item $ \hat{D}_{8}^{(1)}$

$ r_{i}  \in  D_{8}^{(1)}~~~ for ~ r_{i}\in \{1,....,9\}; $
$ ~~ r_{10} =  - K^{+} - K^{-} $

\item $\hat{E}_{8}^{(1)} $

$ r_{i}  \in  E_{8}^{(1)}~~~ for ~ r_{i}\in  \{1,....,9\}; $
$ ~~ r_{10}  =  - K^{+} - K^{-} $

\end{enumerate}

\pagebreak
\setcounter{page}{38}

FIGURES CAPTIONS

\begin{description}

\item[Fig. 1] The first states of the first three levels of $ (1,0) $
representation of the BA subalgebra in the $ (0,1,0) $ of the Hyp-KMA
$\hat{A}_{1}^{(1)}$ are reported.

A state is denoted by a dot, the full and dotted arrows denote resp. the action
 of the BA and Hyp-KMA generators.

Vertical (horizontal) full arrow corresponds, respectively, to the action of
$ E^{-\hat{\alpha}_{1}}$ ($ E^{-\hat{\alpha}_{2}}$).
Dotted arrow pointing to the right (left) corresponds, respectively, to the
action of $ E^{-\alpha_{2}}$ ($ E^{-\alpha_{1}}$).

The integers in brackets gives the weight of the states respect the BA
subalgebra.

\end{description}

\pagebreak

\end{document}